\documentclass[12pt,preprint]{aastex}

\newcommand\nult{{\nu_{\rm LT}}}
\newcommand\Eavg{{E_{\rm avg}}}
\newcommand\dE{{\Delta E}}
\newcommand\ieff{{i_{\rm eff}}}
\newcommand\iBH{{i_{\rm BH}}}

\shorttitle{Ring Model for LFQPOs}
\shortauthors{Schnittman, Homan, \& Miller}

\begin{document}

\title{A Precessing Ring Model for Low-Frequency Quasi-periodic
Oscillations}
\author{Jeremy D.\ Schnittman\altaffilmark{1,~}\altaffilmark{2}, Jeroen
Homan\altaffilmark{2}, and Jon M.\ Miller\altaffilmark{3}}
\altaffiltext{1}{Department of Physics, University of Maryland,
82 Regents Drive, College Park, MD 20742, {\tt schnittm@umd.edu}}
\altaffiltext{2}{Kavli Institute for Astrophysics and
Space Research,
Massachusetts Institute of Technology, 77 Massachusetts Avenue,
Cambridge, MA 02139, {\tt jeroen@space.mit.edu}}
\altaffiltext{3}{Department of Astronomy, University of Michigan, 500
  Church Street, Dennison 814, Ann Arbor, MI 48109, {\tt
    jonmm@umich.edu}} 
%\maketitle

\begin{abstract}
We develop a simple physical model to describe the most common type of
low-frequency quasi-periodic oscillations (QPOs) seen in a number of
accreting black hole systems, as well as the shape of the
relativistically broadened iron emission lines that often appear
simultaneously in such sources. The
model is based on an inclined ring of hot gas that orbits
the black hole along geodesic trajectories. For spinning black
holes, this ring will precess around the spin axis of the
black hole at the Lense-Thirring (``frame-dragging'') frequency. Using
a relativistic ray-tracing code, we calculate X-ray light curves and
observed energy spectra as a function of the radius
and tilt angle of the ring, the spin magnitude, and the inclination of
the black hole. The model predicts higher-amplitude
QPOs for systems with high inclinations, as seen in a growing
number of black hole binary systems. We find that the {\it Rossi X-ray
Timing Explorer}
observations of low-frequency QPOs in GRS 1915+105 are consistent with
a ring of radius $R \approx 10M$ orbiting a black hole with spin $a/M
\approx 0.5$ and inclination angle of $\iBH\approx 70^\circ$. Finally, we
describe how future X-ray
missions may be able to use simultaneous timing and spectroscopic
observations to measure the black hole spin and probe the inner-most
regions of the accretion disk.
\end{abstract}

\keywords{black hole physics -- accretion disks -- X-rays:binaries}

\section{INTRODUCTION}\label{intro}
Over the past decade, X-ray timing and spectroscopy observations of
accreting black hole binaries have provided us with increasingly sensitive
measurements of the
inner-most regions of the accretion disk. With the proper
theoretical understanding of these observations, we should be able to
probe the behavior of matter and energy in the strongest known
gravitational fields. Recent observations by \citet{mille05} with the
{\it Rossi X-ray Timing Explorer} ({\it RXTE}) provide
time-varying measurements of spectroscopic 
features for a black hole on sub-second time scales. They find a strong 
correlation between the phase of the light curve oscillation and the
width of the iron K$\alpha$ emission line.

We propose a simple model to explain this correlation
and make a number of additional predictions for the spectral features
and light curves of similar black hole binary systems. The quasi-periodic
oscillations (QPOs) considered in this paper
are so-called ``type C'' QPOs, following the naming convention
described in \citet{remil02}.
These appear to be the most common type of QPO in black hole binaries;
they are seen in the spectrally hardest states, are stronger and more
coherent than types A and B, and are observed over a large range in
frequency (typically between $\sim 0.1$ Hz and $\sim 10$ Hz), even
within a single source. It is
believed that this type of black hole QPO is related to the 10-60 Hz QPOs
observed in many neutron star low-mass X-ray binaries
\citep{wijna99,casel05}. 

\citet{stell98} suggested for these
10--60 Hz neutron star QPOs that their frequencies might
correspond to the Lense-Thirring frequency of test particles around a
spinning neutron star. A later version of their model
was also applied to low-frequency QPOs in black hole X-ray binaries
\citep{stell99}. In this work we expand on that model, using a fully
relativistic ray-tracing model, and for the first time compute
quantitative X-ray light curves and emission line spectra that may be
compared more directly with observations. This is similar to the
ray-tracing analysis described in \citet{schni04} for the hot spot
model of high frequency QPOs. However, that hot spot model,
which was developed primarily to explain the 3:2 ratio in the
frequencies of QPO peaks, requires that the emission comes from a
specific radius in the disk where the coordinate frequencies have
small integer ratios \citep{abram01,abram03}. The precessing ring
model presented in this paper requires no such special radius,
consistent with observations, where the LFQPOs are seen to
drift in frequency (and thus radius), while the pairs of high
frequency QPOs are always seen at the same frequency within a given
source \citep{mccli05}.

One of the primary objections to the interpretation of QPOs in the
context of gravitomagnetic frame-dragging is based on the
Bardeen-Petterson effect \citep{barde75}, which predicts that the
inner regions ($r \lesssim 100 M$) of a viscous accretion disk should
align perpendicular to the black hole spin axis. If the disk is
constrained to the black hole equatorial plane, then the
Lense-Thirring precession will not produce a time-varying light curve,
as the observed orientation of the disk does not change. However,
subsequent theoretical studies of the inner accretion disk have led to
a number of
different scenarios in which the emitting gas could have a significant
inclination relative to the black hole spin axis. These perturbations
could be driven by trapped pressure gradients \citep{kato91,ipser96},
radiation warping \citep{iping90,pring96,malon96,malon97}, or
gravitomagnetic forces \citep{marko98}. \citet{marko98} argue that the
radiation warping modes will be strongly damped, while the
lowest-order gravitomagnetic modes, corresponding to
the Lense-Thirring precession frequency, will be weakly damped and
should exhibit quality factors of $Q \sim 2-50$, in agreement with
observations. 

In the context of the
supermassive black hole in Sgr A$^\ast$, \citet{rocke05} have recently
shown that for hot, magnetized disks with low Mach-numbers,
internal pressure gradients can sustain large-scale, coherent
precession of the entire inner-disk region. This picture of a
hot, geometrically thick region in the inner disk is consistent
with some of the earliest models of black hole accretion disks
\citep{thorn75}, as well as more recent general relativistic
magneto-hydrodynamic 
simulations that predict a local density maximum, or ``inner torus''
just outside the inner-most stable circular orbit
\citep{devil03a,devil03b}. Additionally, the fact that most black hole QPOs
appear predominately in the power-law dominant spectral states and
usually show greater amplitudes at higher energies further suggests that they
are originating from a hot, non-thermal region of the accretion flow. 
For the purposes of this paper, we are not concerned with the
exact physical mechanism that produces the precessing ring, but rather
we are interested in the observational manifestation of such a ring,
assuming that it exists. 

In Section \ref{model} we describe in greater
detail the features of the precessing ring model and the global disk
geometry. We also show the relation between black hole spin
and ring radius for a given QPO frequency. Section \ref{results}
presents the results of our ray-tracing calculation, giving light
curves and iron line widths for a range of model parameters. In
Section \ref{data} we compare these results to a number of black hole
systems with type C QPOs, in particular the GRS 1915+105 observations
of \citet{mille05}. In Section \ref{discussion} we conclude with a
discussion of how this model may be tested further with future X-ray
missions and observations of active galactic nuclei (AGN).

\section{DESCRIPTION OF THE MODEL}\label{model} For the basic
geometry of this model, we consider a circular ring made up of
massive test particles on geodesic orbits around a Kerr black hole.
This ring can be thought of as a stream of hot gas with finite
thickness, emitting X-rays isotropically in the {\it RXTE/PCA}
energy band ($\sim 2-60$ keV). We assume that the majority of
the observed high-energy flux, including the broadened iron emission
line, is produced by this hot ring. In order to produce a fluorescent
line from non-ionized iron, it may be best to think of the ring as a
relatively cool thermal emitter, surrounded by a hot corona that
produces the power-law flux seen in observations. However, we
note that a hot ring may well have multiple ionization states, which
could complicate the analysis of the broadened iron emission line.

Unlike a number of theoretical models for the integer ratios seen in
high-frequency QPOs, which require specific geodesic frequencies,
here there is no special radius required for
the ring to form. In other words, given the black hole mass and spin,
a ring of a specific radius will precess at one particular frequency,
but there is nothing physically special about that radius or
frequency, consistent with the fact that the low-frequency QPOs tend
to drift in frequency from one observation to the next. Yet just like
the high-frequency QPO models, the ring model will be sensitive to viewer
inclination angle $\iBH$ (measured from the black hole spin axis),
preferentially selecting high (edge-on) inclinations. 

\begin{figure}[t]
\caption{\label{plotone} Schematic diagram of the inclined ring
  geometry. The ring has a major radius $R$, cross-sectional radius
  $\delta R$, and is inclined at an angle $\Delta\theta$ with respect to the
  black hole equator. The observer is located at an inclination $\iBH$
  with respect to the black hole spin axis, giving an effective
  inclination to the ring axis of $\ieff = \iBH\pm\Delta\theta$.}
\begin{center}
\scalebox{0.8}{\includegraphics{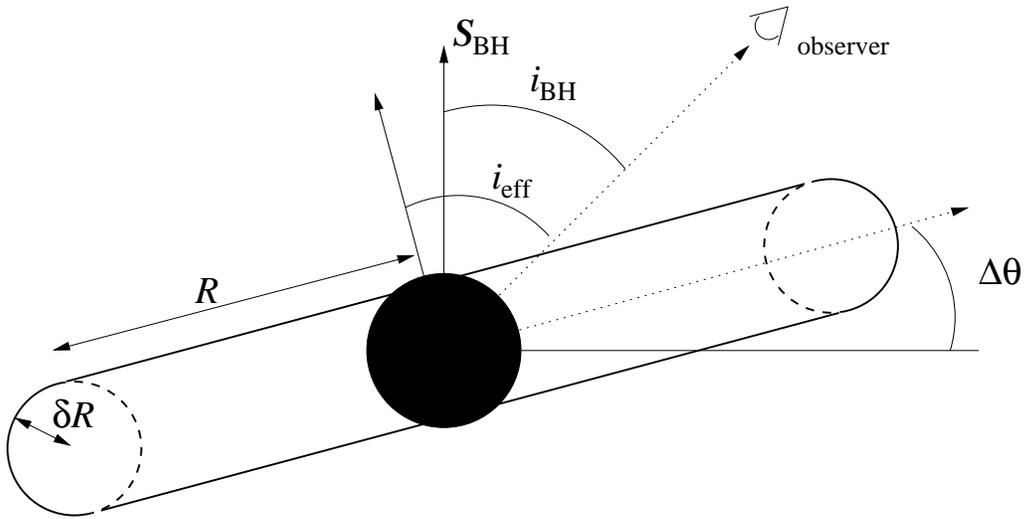}}
\end{center}
\end{figure}

The ring axis is inclined at some angle
$\Delta\theta$ to the black hole spin equator (see Fig.\
\ref{plotone} for a schematic diagram of the ring geometry), and thus
undergoes
Lense-Thirring precession due to the dragging of inertial frames
around the Kerr black hole \citep{barde75}. This misalignment could be
caused by one of two basic mechanisms: (1) the binary orbit of the
donor star and the outer accretion disk
are aligned with the black hole spin axis ($\iBH = i_{\rm bin}$), and
some instability (such as those mentioned in the Introduction)
causes the inner disk to be perturbed and form a ring inclined with
respect to the orbital plane; or (2) the entire binary orbit is
misaligned with the black hole spin ($\iBH = i_{\rm bin}\pm
\Delta\theta$), so that the inner accretion flow is forced to precess
around the spin axis. In either case, the precessing ring will
appear to a distant observer with an inclination angle that varies
periodically, thus modulating the observed X-ray light curve. 

The Lense-Thirring precession frequency $\nult$ is a function of the
black hole mass $M$, spin $a$, and the
radius $R$ of the ring. Specifically, $\nult$ is the difference
between the vertical and azimuthal coordinate frequencies, which for
nearly circular prograde orbits can be written as
\citep{barde72,perez97,merlo99} 
\begin{equation}
\nult = \nu_\phi-\nu_\theta =
\frac{1}{2\pi}\left(\frac{1}{R^{3/2}+a}\right)
\left[1-\left(1-\frac{4a}{R^{3/2}}+\frac{3a^2}{R^2}\right)\right]^{1/2}, 
\end{equation}
where we have adopted geometrized units with $G=c=M=1$. To convert to
cgs units, simply multiply $\nult$ by $2.04\times 10^5
(M/M_\odot)^{-1}$ Hz. 

\begin{figure}[t]
\caption{\label{plottwo} Orbital radius ({\it solid lines}) at which
the Lense-Thirring precessional frequency is equal to the frequency of
the QPO peaks for GRS 1915+105, located near 1 and 2 Hz ({\it thick
and thin lines, respectively}), as a function
of black hole spin $a/M$ (assuming a mass of 
$14.4 M_\odot$). We only consider cases where this radius is outside
of the inner-most stable circular orbit ($R_{\rm ISCO}$), shown here
by a dotted line. Also plotted ({\it dashed lines}) are the
corresponding Keplerian frequencies $\nu_\phi$ for circular geodesic
orbits at the given radius.}
\begin{center}
\scalebox{0.8}{\includegraphics{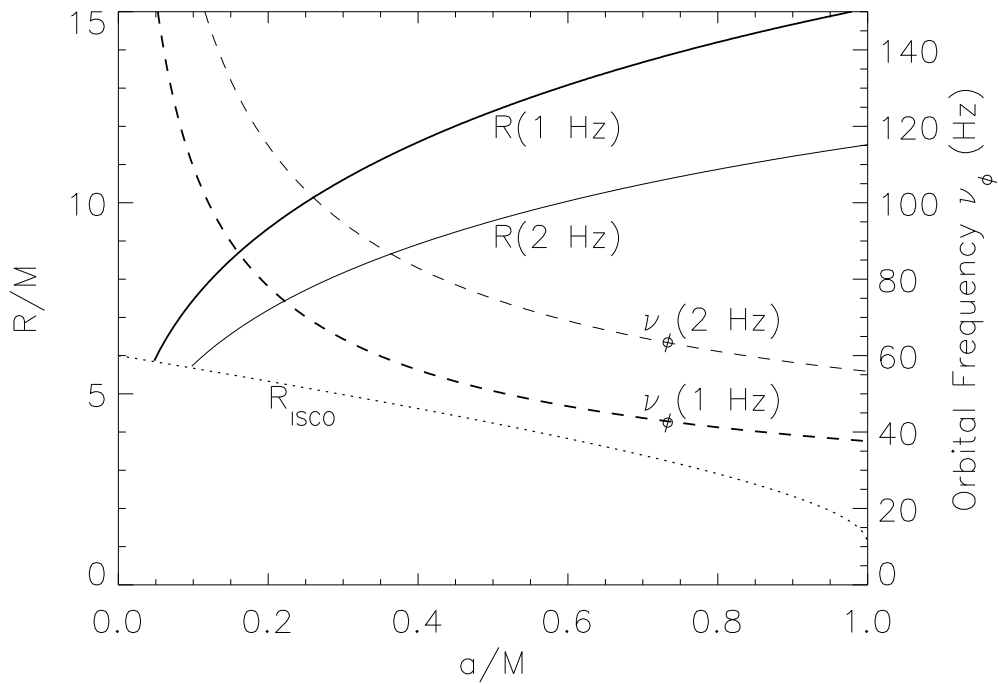}}
\end{center}
\end{figure}

Using the compact binary system of GRS 1915+105 as an example, in
Figure \ref{plottwo} we show how the radius of the ring is a
function of black hole spin and measured QPO frequency. Fixing the
black hole mass at $14.4 M_\odot$ \citep{harla04}, we
plot the radius at which a ring would have a given
precession frequency $\nult$, as a function of black hole
spin. For example, a spin of $a/M=0.5$ would give
$\nult = 1$ Hz at a radius of $R = 12.5M$ and $\nult = 2$ Hz at
$R=9.5M$. Also plotted are the corresponding orbital
frequencies $\nu_\phi$ for geodesic orbits at the given radii. For
smaller spins, the ring must be closer to the black hole to experience
the same frame-dragging effects, and thus it has a higher orbital
frequency. In the limit of $a/M \to 0$, the required radius
moves inside that of the inner-most stable circular orbit (ISCO), and
is not considered physical. From this argument alone, the
precessing ring model gives a modest lower limit on the black hole spin
of $a/M \gtrsim 0.1$ for GRS 1915+105. For ``typical'' black hole
binaries with $M \approx 10M_\odot$ and $\nu({\rm LFQPO}) \approx 10$
Hz \citep{mccli05}, this lower limit would be $a/M \gtrsim 0.25$. 

The entire model can thus be described completely with a few basic
parameters: the black hole mass $M$, spin $a$, and inclination angle
$\iBH$, the ring's radius $R$, and the tilt angle $\Delta\theta$. The
characteristic thickness (cross-sectional radius) of the ring is taken
as $\delta R/R = 0.1$, but for
moderately thick rings ($\delta R/R \lesssim 1/3$), we
find the results to be largely independent of this parameter. As
shown in Figure \ref{plottwo}, for a given black hole mass and QPO
frequency, the spin can be determined by measuring the radius $R$. In
the next Section we show how the light curves and iron emission lines
can be used to infer $R$, as well as the angles $i$ and
$\Delta\theta$.

As in the hot spot QPO model described in \citet{schni05}, the width
of the QPO peaks is caused by the finite lifetime of coherent rings,
as they are continually formed and destroyed and formed again with
random phases. This leads naturally to a Lorentzian peak in the power
spectrum with a full-width half-maximum inversely proportional to the
characteristic coherence lifetimes of the inclined rings. One of the
predictions of the finite-lifetime model is that the higher harmonic
peaks should have the same widths as the fundamental \citep{schni05},
in agreement with observations of GRS 1915+105 \citep{morga97,
mille05}. With this 
model, the observed $Q$-values of $\sim 6-8$ correspond to typical
ring lifetimes of $\sim 2-3$ precessional periods. 

\section{LIGHT CURVES AND IRON LINE SPECTRA}\label{results}

The simulated light curves of the precessing ring were created using a
fully relativistic ray-tracing code for the Kerr metric
\citep{schni04}. As with the hot spots and arcs described in that
paper, here we consider primarily an
optically thin line emission model, where the gas is an isotropic,
monochromatic emitter in the local rest frame of the ``guiding
center'' geodesic orbits. Future work will investigate the effects of
other, more sophisticated emission and absorption models
\citep{schni05b}.  

Even though the effective inclination of the ring will change purely
sinusoidally at a single frequency $\nult$, the light curve can have
significant power at higher harmonics due to relativistic beaming and
gravitational lensing effects \citep{schni04}. These higher harmonics
are clearly seen in the power spectra of many type C QPOs, typically
weaker in power by about an order of magnitude
\citep{morga97,mille05}, consistent with the light curves calculated
below for black hole inclinations of $60-70^\circ$. 

\begin{figure}[tp]
\caption{\label{plotthree} Ray-traced images of the precessing ring model,
shown at various phases of the Lense-Thirring period. Also shown is
the integrated X-ray light curve ({\it lower left}) and the broadened
emission line spectrum ({\it lower right}) for each frame. The black
hole has a spin of $a/M=0.5$,
inclination of $\iBH=70^\circ$, and the ring tilt angle is $\Delta \theta
= 20^\circ$, giving effective inclinations of $\ieff=50^\circ,
70^\circ, 90^\circ$, and $70^\circ$ in the four frames shown. This
geometry is characterized
by three distinct peaks in the spectrum, due to the photons beamed
away from the observer on the left, towards the observer on the right,
and a middle peak from the gravitational magnification of gas moving
transverse relative to the observer.}
\begin{center}
\scalebox{0.5}{\includegraphics*[10,400][425,750]{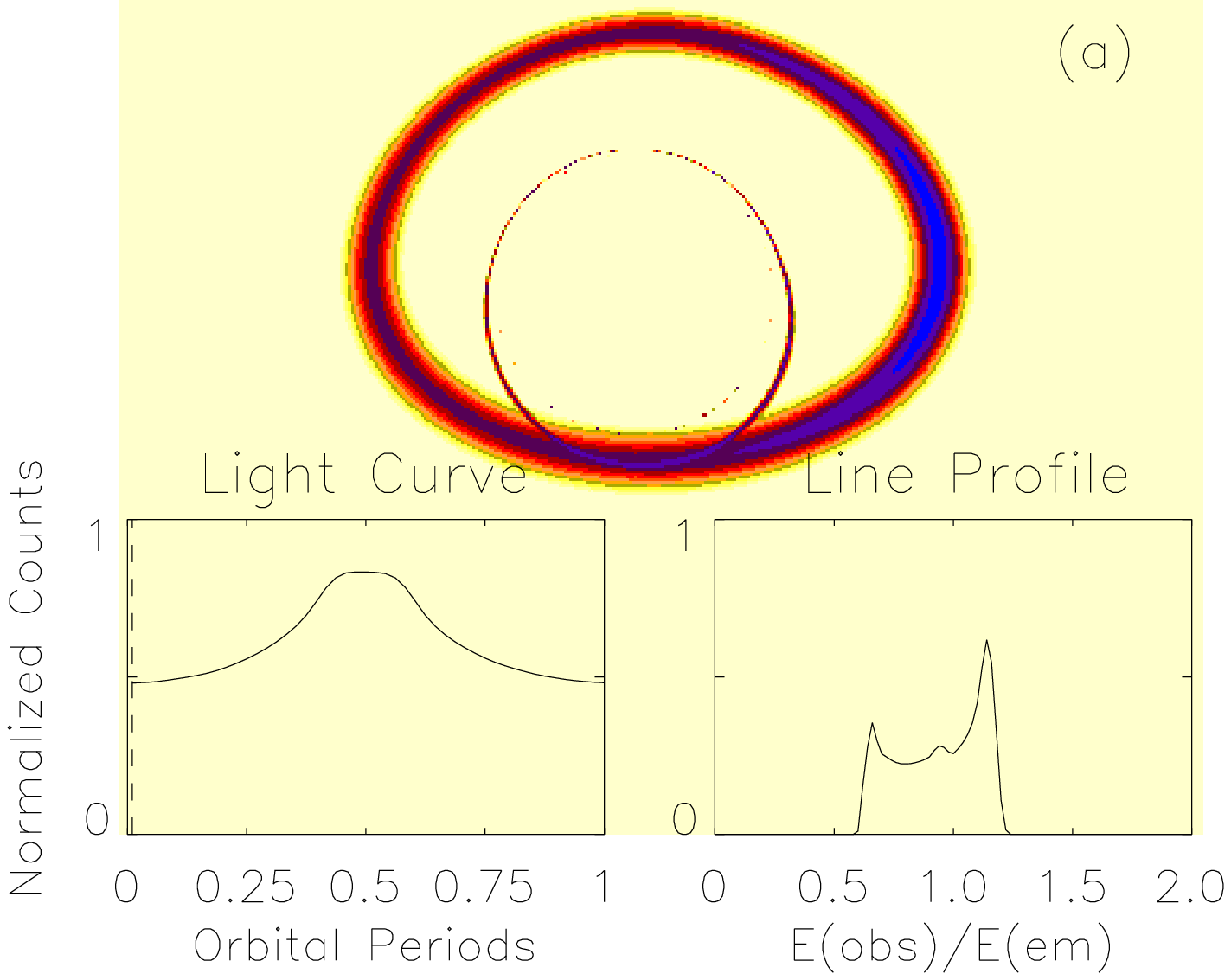}}
\scalebox{0.5}{\includegraphics*[40,400][425,750]{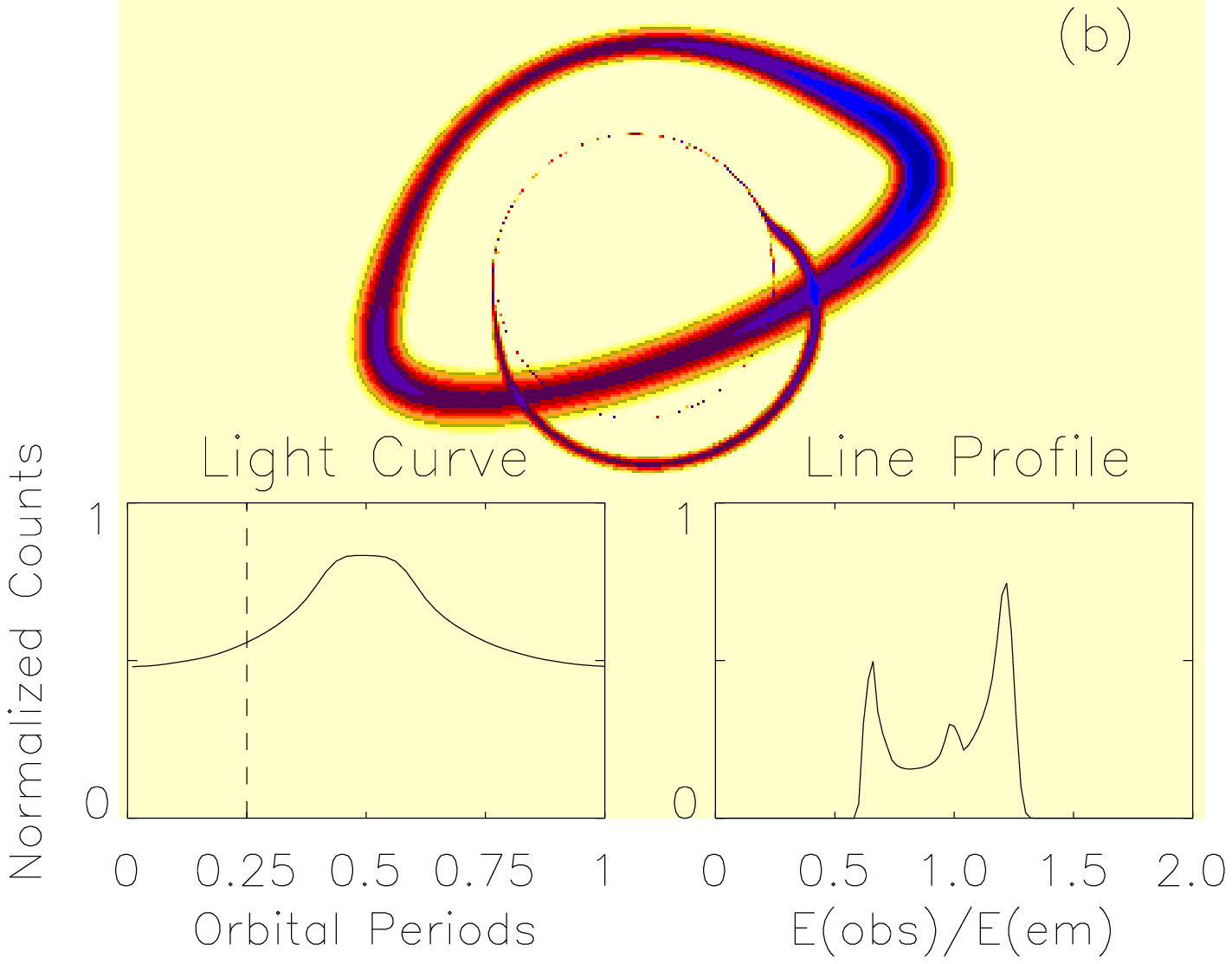}}
\scalebox{0.5}{\includegraphics*[40,400][425,750]{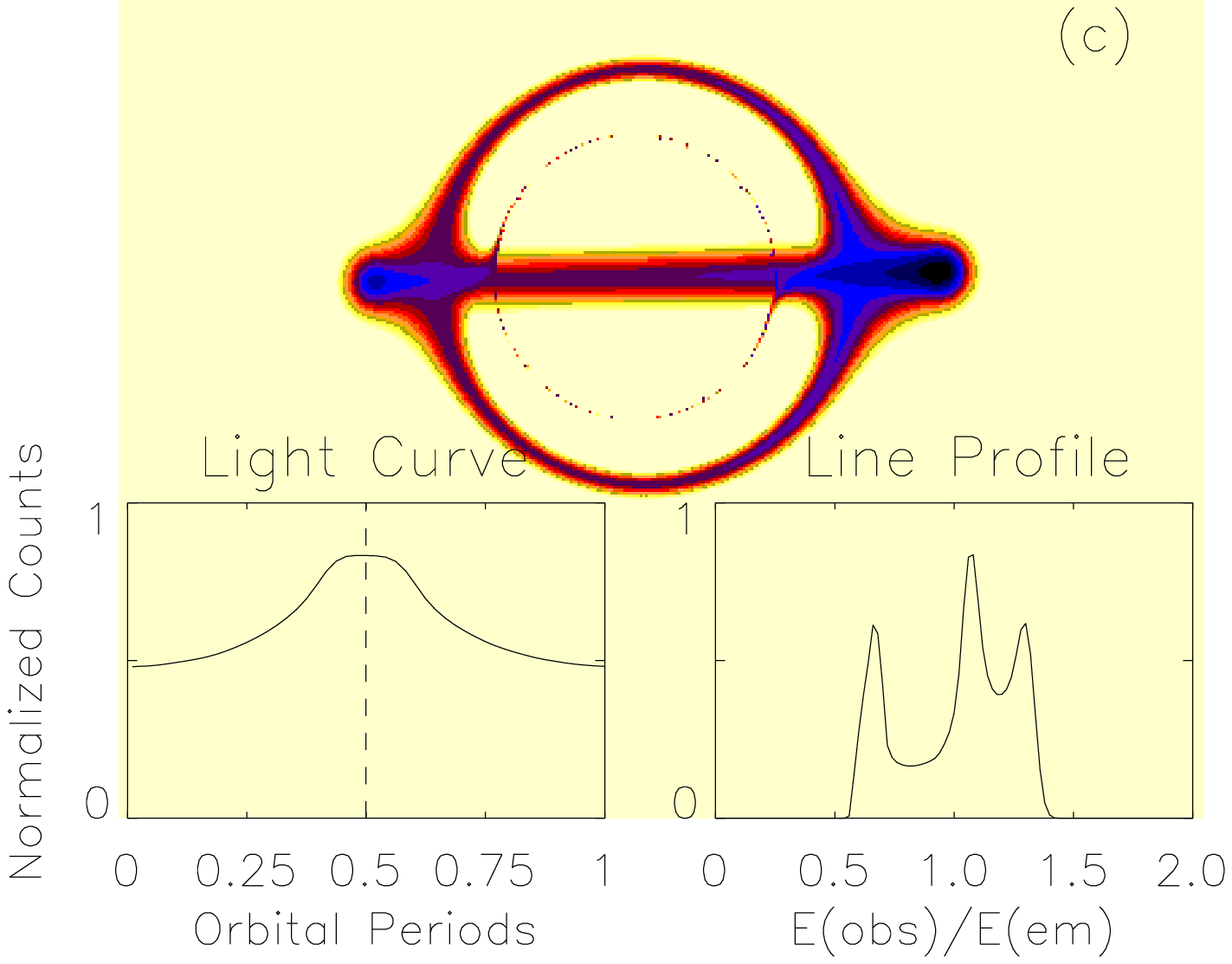}}
\scalebox{0.5}{\includegraphics*[40,400][425,750]{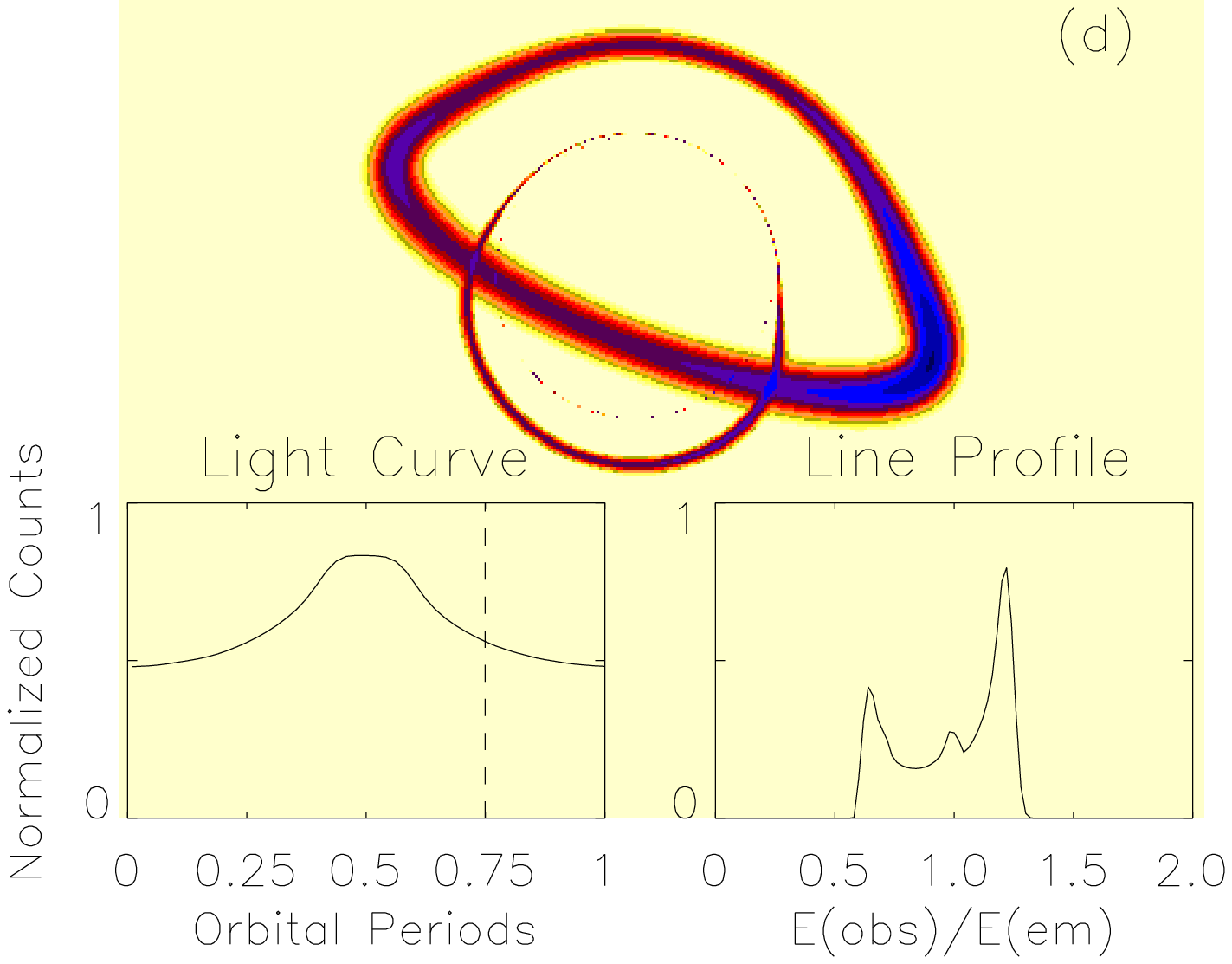}}
\end{center}
\end{figure}

In Figure \ref{plotthree} we show a series of snap-shots of how the
precessing ring
would appear to a distant observer, for a representative spin of
$a/M=0.5$ and radius $R=9.5M$. The black hole inclination is
$\iBH=70^\circ$ and the tilt angle is $\Delta\theta=20^\circ$. The images
are color-coded by relative intensity on a logarithmic scale. The gas
orbiting the black hole is moving towards the observer on the right
side of each image, and moving away on the left side, respectively
giving the blue-shifted and red-shifted spectral peaks plotted in the
lower right of each frame. For both the light curves (plotted in the
lower left of each frame) and energy spectra, the
vertical axis is normalized intensity in units of [\#
photons/s/cm$^2$].

At the peak of the light curve (frame ``c'' in Fig.\ \ref{plotthree}),
the effective inclination is $\ieff = \iBH+\Delta\theta=90^\circ$,
resulting in a
flux maximum due to both special relativistic beaming and the
formation of an Einstein ring through the gravitational lensing of the
emission from the far side of the black hole. This lensing
produces a third, intermediate spectral peak from the gravitational
magnification of gas moving transverse to the observer's line of
sight \citep{beckw04,schni05b}. While this
third peak is most pronounced at very high inclinations where lensing is
more important, it should also be observable for $\ieff\approx 70^\circ$
(frames ``b'' and ``d'') with a next-generation
X-ray spectroscopy mission.

Due in large part to these relativistic effects, the amplitudes of
the observed QPOs are primarily sensitive to the angles $\iBH$ and
$\Delta\theta$. The effective inclination as a function of time can be
written 
\begin{equation}\label{i_eff}
\ieff(t) = \iBH + \Delta\theta \cos(\nult t + \psi),
\end{equation}
where $\psi$ is some arbitrary phase. Understandably, the greater the
range of $\ieff$, the greater the QPO amplitude. However,
because of the planar symmetry around $\ieff=90^\circ$,
which maps $\ieff \to (180^\circ-\ieff)$, systems with
very high $\iBH$ actually produce smaller light curve modulations for the
same $\Delta\theta$. Additionally, for edge-on systems with
$\iBH=90^\circ$, symmetry dictates that
the light curve is modulated primarily at frequency $2\nult$, with no
power at the fundamental $\nult$ \citep{bursa04}.

Figure \ref{plotfour} illustrates this dependence by plotting the
simulated rms amplitude as a function of black hole
inclination for a range of tilt angles $\Delta\theta$. We set fixed
the spin $a/M=0.5$ and ring radius $R=9.5M$, corresponding to
$\nult=2$ Hz for $M=14.4M_\odot$. For a given tilt angle
$\Delta\theta$, we see that the greatest amplitude fluctuations occur
when
$\iBH+\Delta\theta=90^\circ$. For example, if $\iBH=80^\circ$ and $\Delta
\theta = 20^\circ$, then the effective range of $\ieff$ is
only $[60^\circ .. 90^\circ]$, while for $\iBH=70^\circ$ the range
extends to $[50^\circ .. 90^\circ]$, giving a larger light curve
modulation. We find that the rms amplitude is relatively insensitive
to the black hole spin, being slightly larger for smaller
values of the spin parameter and thus ring radius. In the paradigm
where the binary orbit is aligned with the black hole spin ($\iBH =
i_{\rm bin}$), a thick outer accretion disk may obscure some of the
emission for $i_{\rm bin} \gtrsim 80^\circ$, limiting the potential
range of observations \citep{naray05}. Yet if the outer disk is 
misaligned with the black hole axis (see below, Section \ref{data}),
the ring could be unobscured even for very large values of $\iBH$.

As can be seen clearly from the frames of Figure \ref{plotthree}, the
effective inclination affects not only the total observed intensity, but
also significantly changes the shape of the relativistic emission
line. With our current level of spectroscopic and timing sensitivity,
the detailed features of these spectra can not be discerned on the QPO
time scale. Yet, as shown in \citet{mille05}, the emission
line width and centroid may be measurable with sufficient accuracy to
constrain the precessing ring model. 

\begin{figure}[t]
\caption{\label{plotfive} (a) Center of relativistically shifted iron
emission line $E_{\rm avg}$ as a function of black hole spin, for an
inclination angle of $\iBH=70^\circ$ and radius of $R=9.5M$. (b) The line
widths $\Delta E$ for the same
broadened emission lines. For both plots, the spectra are taken at the
phases corresponding to the light curve minima ({\it below dashed curve})
and maxima ({\it above dashed curve}).}
\begin{center}
\scalebox{0.7}{\includegraphics{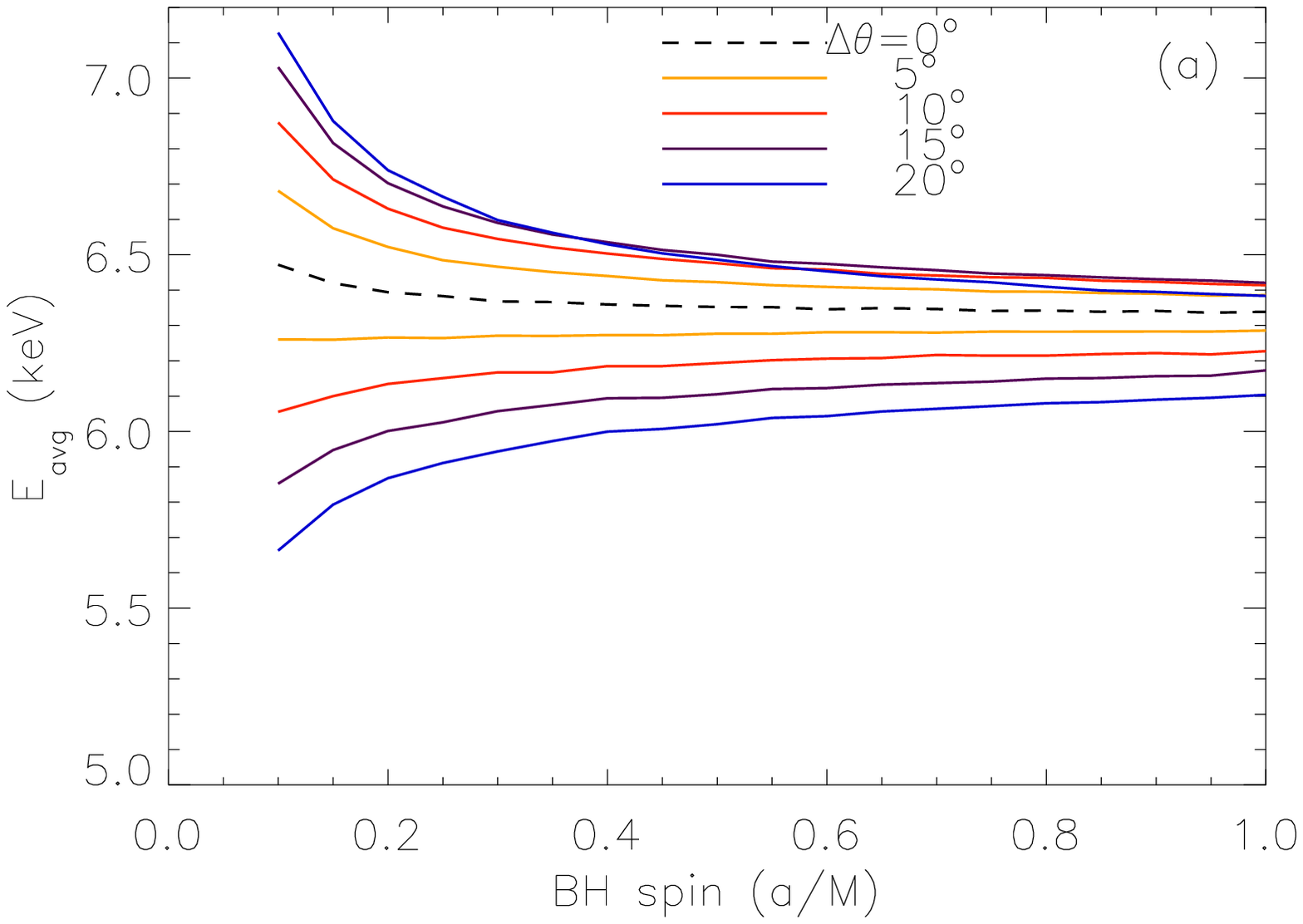}}
\scalebox{0.7}{\includegraphics{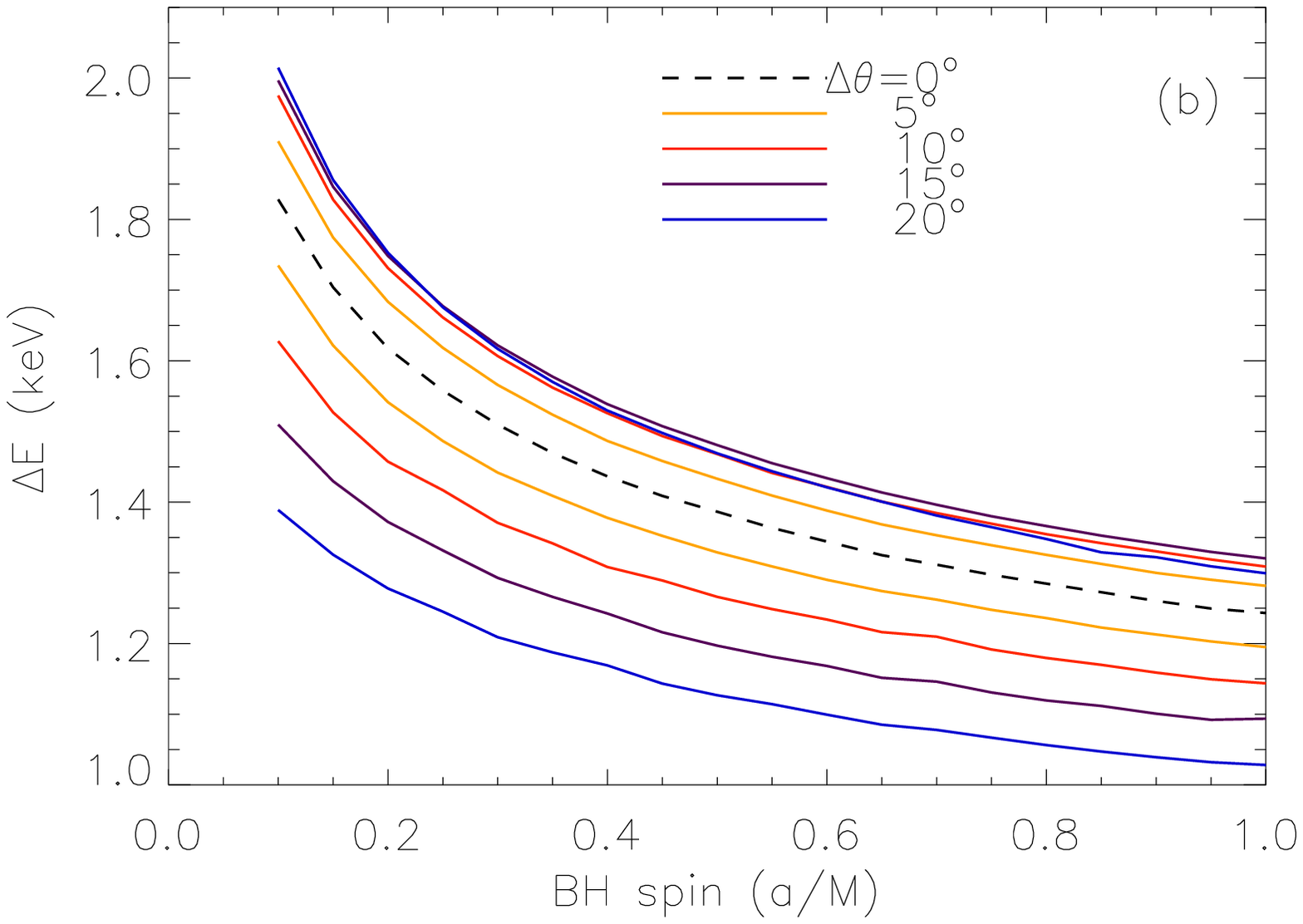}}
\end{center}
\end{figure}

Following that approach, in Figure \ref{plotfive} we plot the line
centroid $\Eavg$ and width (standard deviation) $\dE$ as a function of
black hole spin, for a fixed inclination of $\iBH=70^\circ$ and QPO
frequency of 2 Hz.
In order to best understand these results, recall
Figure \ref{plottwo}, which shows the dependence of
orbital frequency $\nu_\phi$ on spin. For a given QPO frequency
$\nult$, smaller values of the spin correspond to smaller radius rings
and thus larger orbital velocities. As $\nu_\phi$ increases, so does
the relativistic Doppler broadening, as the emitting gas is moving
faster, both towards and away from the observer. 

While the gravitational red-shift at smaller radii reduces the energy
of all emitted photons equally, giving an observed $\Eavg$ lower than
the rest frame energy, special relativistic beaming causes the observer to see
a larger number of blue-shifted photons (through the invariance of
$I_\nu/\nu^3$), thus {\it increasing} $\Eavg$. 
At high inclinations ($\ieff \gtrsim 70^\circ$) the beaming factor
dominates, giving larger values of $\Eavg$ at smaller radii (spin
value), while at lower inclinations ($\ieff \lesssim 70^\circ$) the
gravitational red-shift wins out and reduces $\Eavg$ for smaller
spins. The upper curves (above the dashed line) in Figure
\ref{plotfive}a are taken at the light
curve maximum, corresponding to the highest effective inclination,
while the lower curves have successively smaller values of $\ieff$
and thus smaller $\Eavg$. 

The width of the emission line $\dE$, shown in Figure \ref{plotfive}b,
is a direct measurement of the range of Doppler shifts seen by the
observer. As the effective inclination increases, the gas moves
towards and away from the observer with higher velocity, broadening
the line more. However, in the limit of $\ieff \to 90^\circ$, while
the greater beaming helps broaden the line, the lensing effects
actually serve to {\it decrease} the width $\dE$ as the
gravitational magnification amplifies the contribution of a small
region of the ring, giving the central third peak of the iron line
spectrum (see Fig.\ \ref{plotthree}c). Therefore
the line properties $\Eavg$ and $\dE$ are nearly identical for
$\ieff \approx 80^\circ-90^\circ$, as can be seen in the
top three curves in Figures \ref{plotfive}a and \ref{plotfive}b. Of
course, higher resolution observations of these lines could break the
inclination degeneracy by giving detailed spectral features other than
just $\Eavg$ and $\dE$.

\begin{figure}[t]
\caption{\label{plotsix} (a) $E_{\rm avg}$ and (b) $\Delta E$ as in
  Figure \ref{plotfive}, but for a black hole inclination of
  $\iBH=30^\circ$. For both plots, the spectra are taken at the
  phases corresponding to the light curve minima ({\it below dashed curve})
  and maxima ({\it above dashed curve}). Note that the top curves
  in each of these plots correspond directly to the bottom curves in
  Figure \ref{plotfive}: 
  $\ieff({\rm min}) = \iBH-\Delta\theta=70^\circ-20^\circ$ and 
  $\ieff({\rm max}) = \iBH+\Delta\theta=30^\circ+20^\circ$. }
\begin{center}
\scalebox{0.7}{\includegraphics{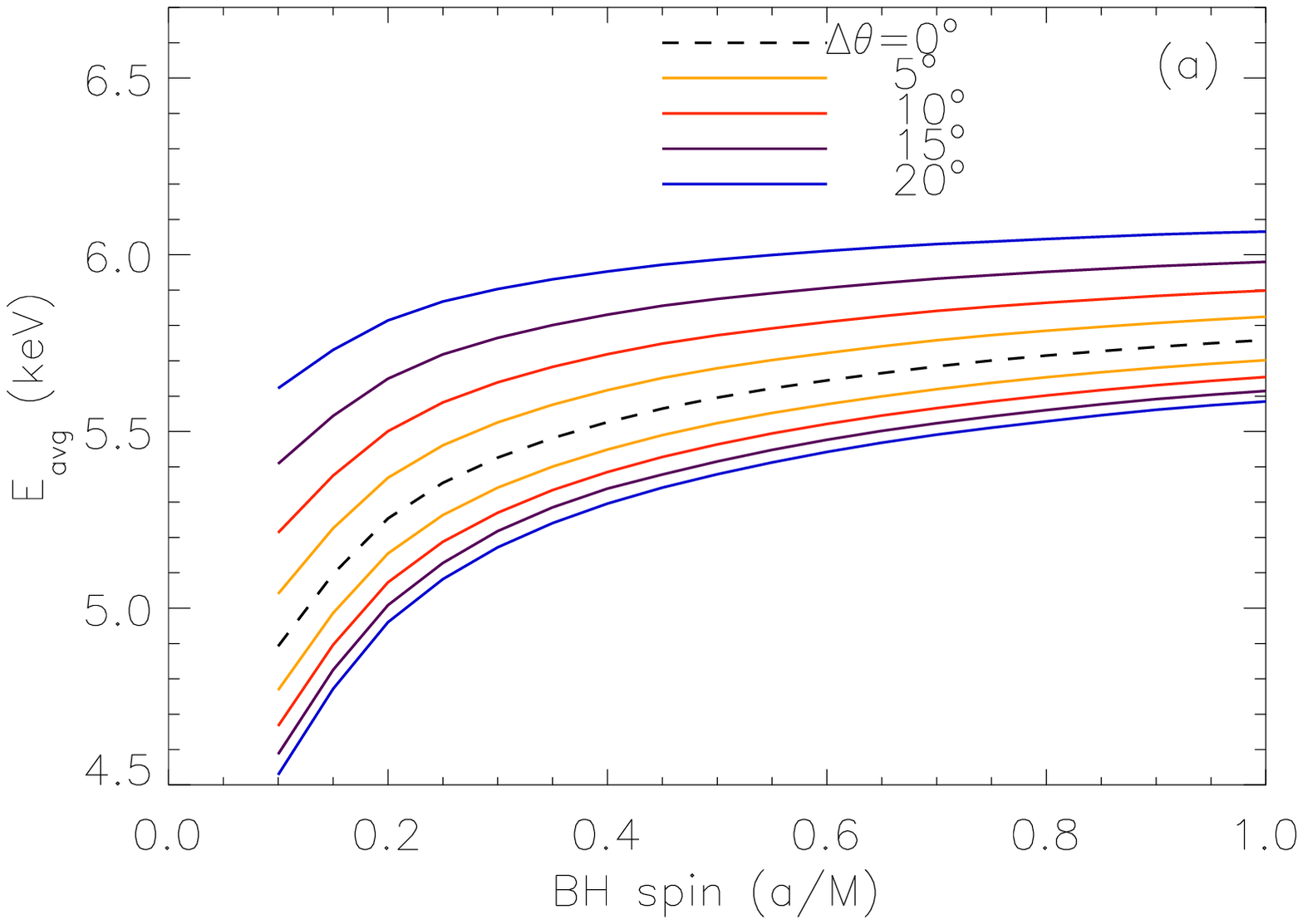}}
\scalebox{0.7}{\includegraphics{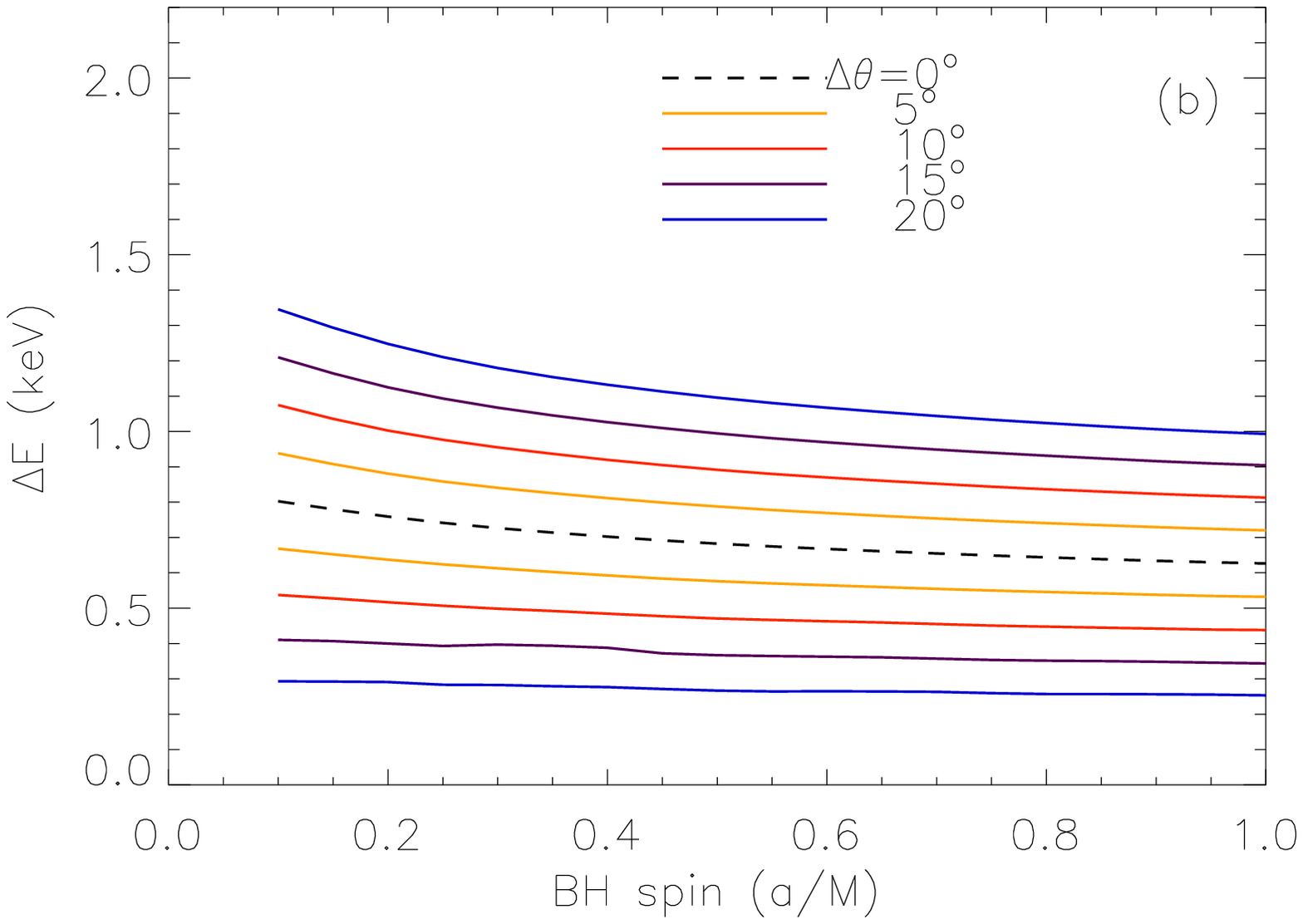}}
\end{center}
\end{figure}

As a point of comparison, we show in Figure \ref{plotsix} the same
iron line predictions for an inclination angle of $\iBH =
30^\circ$. As in Figure \ref{plotfive}, we see that the range of
$\Eavg$ and $\dE$ are greater at small spin values where the orbital
velocities are larger. However, due to the smaller inclination angle,
relativistic beaming plays a significantly smaller role, so the
gravitational redshift dominates the behavior of $\Eavg$ at small
spin and radius (Fig.\ \ref{plotsix}a). Similarly, the nearly face-on
projection gives a much smaller range of redshift around the ring,
corresponding to smaller values of $\dE$ (Fig.\ \ref{plotsix}b). 

By measuring the iron line profiles at the maxima and minima
of the QPO light curve, we have another method for potentially
determining a black hole's spin. As can be seen in Figures
\ref{plotfive}a and \ref{plotsix}a, larger
values of the spin parameter correspond to a smaller range in $\Eavg$
between maximum and minimum intensity. While the range in $\Eavg$ is
also sensitive to $\Delta\theta$, we can use measurements of $\dE$ to
break this degeneracy: the difference between $\Delta E({\rm max})$
and $\Delta E({\rm min})$ is nearly independent of spin for $a/M
\gtrsim 0.3$, so measuring this difference should give a robust
measurement of $\Delta\theta$. Then with this value for
$\Delta\theta$, we can use the line centroids plotted in Figure
\ref{plotfive}a to determine the spin. However, even this simple
method can be plagued by systematic errors described in the following
Section. 

\section{COMPARISON WITH OBSERVATIONS}\label{data}

\begin{figure}[t]
\caption{\label{plotfour} QPO amplitude of the X-ray light curve
fluctuations as a function of the inclination angle $\iBH$ of
the black hole spin axis. The ring precesses around the spin axis with
a tilt angle of $\Delta \theta$, giving an effective range of
inclinations of $\ieff=\iBH\pm\Delta\theta$. The black hole has spin
$a/M=0.5$ and the ring has radius $R=9.5M$.}
\begin{center}
\scalebox{0.8}{\includegraphics{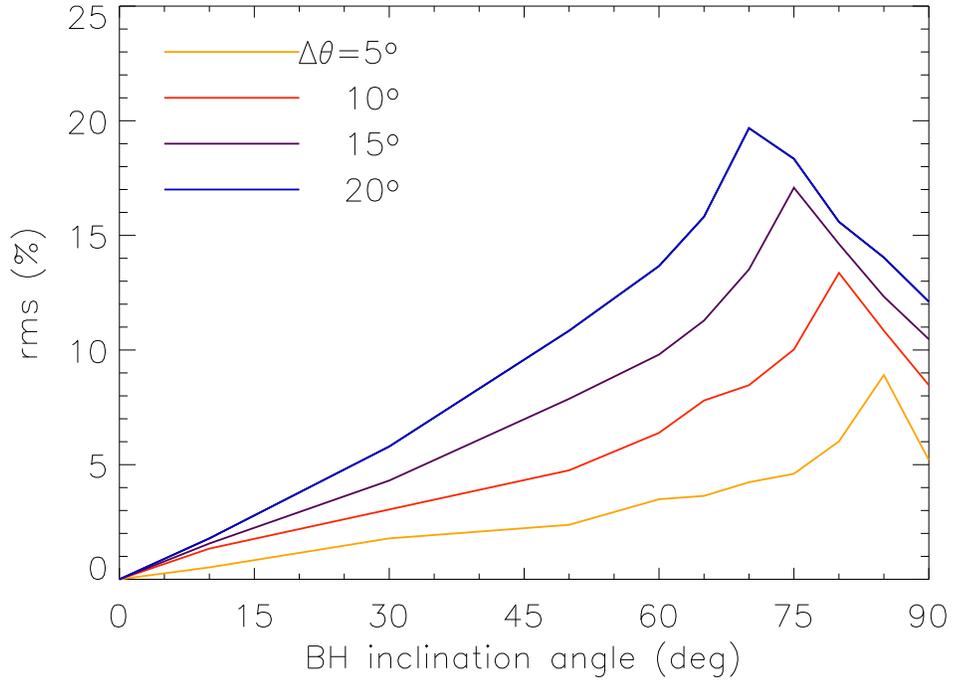}}
\end{center}
\end{figure}

One of the predictions of our model that can be tested with current
instrumentation ({\it RXTE}) is that type C QPOs should in general be
stronger in sources with a higher inclination angle (see Fig.\
\ref{plotfour}). We made a first attempt to test this prediction by
selecting black hole X-ray
transients with well-known binary inclinations that have shown type C QPOs
in {\it RXTE} data. At present, there are only a handful of sources
that meet both criteria, with just two sources (4U 1543--47 and GX
339--4) having relatively well-determined low inclinations. In Table
\ref{tableone} we list the sources included in our survey. We have
included one source, XTE J1650--500, whose
inclination is somewhat uncertain but might fill in the gap between
the high inclination ($i_{\rm bin} \gtrsim 70^\circ$) and low inclination
($i_{\rm bin} \lesssim 30^\circ$) sources. In an effort
to compare similar QPO modes, we have selected {\it RXTE} observations
in which these sources showed type C QPOs that were either close to
1--1.5 Hz
or close to 4.5--5 Hz, since we could not identify a single
frequency that was observed in all sources. Of course, this is quite
reasonable considering the range of black hole masses and (presumably)
spins involved, i.e.\ the same values of $R/M$ correspond to a range
of precession frequencies in different sources. 

\begin{figure}[t]
\caption{\label{plotseven} Dependence of rms amplitude on inclination for the
black hole X-ray binaries listed in Table \ref{tableone}. The figure
shows type C QPOs observed in the frequency ranges 1-1.5 Hz ({\it
squares}) and 4.5-5 Hz ({\it circles}). For XTE J1650--500, arrows
indicate the lower limit on the inclination. As predicted in Figure
\ref{plotfour}, the data shows a clear trend that suggests
higher-inclination systems will have higher-amplitude QPOs.}
\begin{center}
\scalebox{0.8}{\includegraphics[angle=-90]{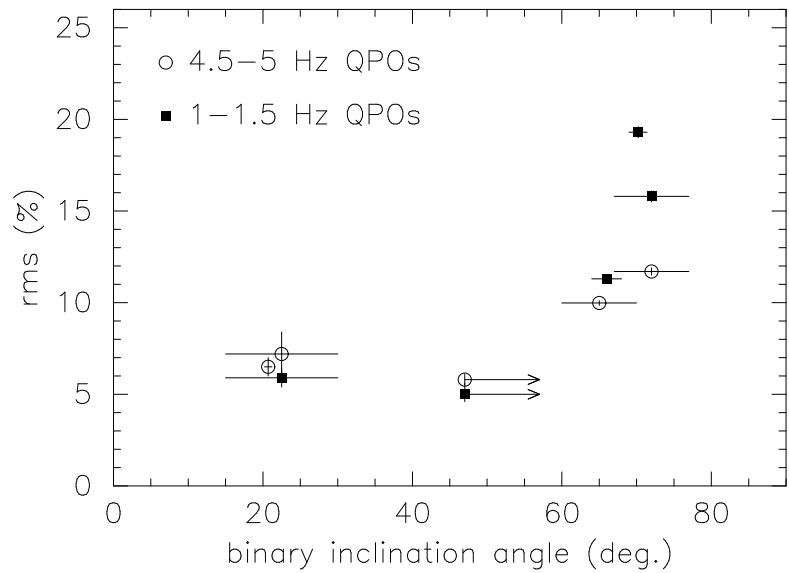}}
\end{center}
\end{figure}

For all of the sources, we constructed Poisson-subtracted power spectra using
data from the entire PCA bandpass (2--60 keV), following methods
described in \citet{homan05}. The power spectra were fit with
combinations of broad and narrow Lorentzians, and the rms values quoted
in Table \ref{tableone} were calculated by integrating the narrow
Lorentzians over frequency from 0 to $\infty$ Hz. The
QPOs had $Q$-values between 4 and 11.5. The quoted QPO frequencies
correspond to the frequency at which the Lorentzian reaches its
maximum  in $\nu P(\nu)$, where $P(\nu)$ is the power in units of
$[{\rm (rms/mean)}^2/{\rm Hz}]$. In order to provide a visual
comparison with the precessing ring model, in Figure \ref{plotseven} we
reproduce the data from Table \ref{tableone}, plotting QPO-rms versus
binary inclination for these seven sources. Although our
dataset is arguably limited, there are clear indications that the rms
amplitudes are indeed higher in the sources with the highest
inclinations, and are roughly in agreement with the results
shown in Figure \ref{plotfour}. In most cases one or more harmonics
were detected, but in the two sources with the lowest inclinations
(4U 1543--47 and GX 339--4), those features were very weak, in
agreement with the model predictions. 

Assuming that the {\it RXTE} light curves come primarily from
the hot ring emission, we can use the QPO amplitudes to constrain
the accretion geometry of the system. Again using GRS 1915+105 as an
example, we take the inclination of the outer disk as the binary
inclination, given by \citet{fende99} as $i_{\rm bin} = 66^\circ\pm
2^\circ$. For the QPO near 2 Hz, the observed rms amplitude is 13.6\%
\citep{mille05}. As
mentioned in Section \ref{model}, there are two basic options for the
global geometry of the system. If the black hole spin axis is aligned
with the binary system's orbital angular momentum, then $\iBH=66^\circ$
and from Figure \ref{plotfour} we infer that $\Delta\theta \approx
17^\circ$. If, on the other hand, the precession is caused by a global
misalignment of the outer disk and the spin axis, then the binary
inclination can be understood as the minimum of the effective
inclination: $i_{\rm bin} = \ieff({\rm
min})=\iBH-\Delta\theta=66^\circ$. In this 
scenario, Figure \ref{plotfour} shows that $\iBH\approx 80-85^\circ$ and
$\Delta\theta \approx 15-20^\circ$ is consistent with an rms of
13.6\%. However, at least for the case of GRS 1915+105, the lower
level of harmonic power ($\sim 10\%$ of the fundamental) seems to
point to the first scenario, with the disk and spin axis aligned,
since a spin inclination of $\iBH \gtrsim 80^\circ$ would produce a light
curve with harmonic power roughly equal to that of the fundamental. 

Constraining the broad iron line predictions of the ring model with
the current {\it RXTE} energy spectra has proven
somewhat more difficult. For GRS 1915+105, \citet{mille05} find that
the iron line is significantly broader at the light curve maximum
compared to the minimum, measuring the FWHM as 2.7 and 1.1 keV,
respectively. This corresponds to our $\Delta E$ values of 1.1 and 0.5
keV, in qualitative but not quantitative agreement with Figure
\ref{plotfive}b. If, however, we were to set $\iBH=46^\circ$ and
$\Delta\theta=20^\circ$, and thus $i_{\rm bin} = \ieff(\rm
max)=66^\circ$, the predicted values of $\Delta E$ would
closely fit the observed values for $a/M \gtrsim 0.5$. Yet this
geometry would give significantly smaller QPO power in the
fundamental, and almost no power in the first harmonic, in
disagreement with the data. 

At the same time, the observed values of $\Eavg$ are actually {\it
lower} for the light curve maximum, with $\Eavg({\rm max})=6.3$ keV
and $\Eavg({\rm min})=6.7$ keV, disagreeing qualitatively with the
predictions of  our model (see Fig.\ \ref{plotfive}a). 
This might be due in part to systematic factors
that complicate direct comparison of theory and observation. There
may very well be significant contributions from other
ionization state of iron in the inner disk, as well as contributions
to the energy spectrum from other elements, which could artificially
increase or 
decrease the location of the line center $\Eavg$, as determined by
fitting a single Gaussian peak. While this other emission could also
serve to artificially broaden the line, including iron K$\alpha$
emission from other, less relativistic regions of the disk would
have the opposite effect of narrowing the observed line, {\it
decreasing} the measured value of $\Delta E$. 

With the quality of present data, we believe that the precessing ring
model should be able to give good constraints on the overall geometry
of the inner accretion region from the light curve amplitudes and
harmonics. On the other hand, the spectral resolution currently
available is not high enough to use the iron lines to constrain the
model parameters or
effectively measure the black hole spin. Perhaps simultaneous
observations could use the timing capabilities of {\it RXTE} to
measure the QPO amplitudes and thus $\iBH$ and $\Delta\theta$, while the
high-resolution spectroscopy of {\it XMM-Newton} or {\it Chandra}
could give more detailed measurements of the iron lines, even if they must
be averaged over the QPO period.

\section{DISCUSSION AND CONCLUSIONS}\label{discussion} We have
developed a geodesic precessing ring model to help explain the
properties of type C low-frequency QPOs in black hole binaries.
In this model, the surrounding accretion disk may be slightly
misaligned  with the black hole spin axis (as would generally be the
case for compact binaries), or the inner-most region of the disk may
get excited into an inclined orbit, forming a ring of hot gas
precessing around the black hole spin axis at the Lense-Thirring
frequency. For a given QPO frequency, we have shown the dependence of
the ring's radius on the black hole spin, which gives a lower limit
on the spin parameter, assuming the ring must be outside of the ISCO.
Using a relativistic ray-tracing code, we have produced X-ray light
curves and time-varying iron line emission spectra. Similar to the
hot spot model described in \citet{schni04}, the ring model predicts
the QPO amplitudes should increase with increasing inclination, in
agreement with observations. Another important feature predicted by
this model is the existence of a third, intermediate peak in the
broad iron line spectrum at the QPO phase corresponding to maximum
intensity, formed by the gravitational lensing of emission from the
far side of the black hole.

In the states where most of the type C QPOs are observed, the
emission of black hole X-ray binaries in the {\it RXTE} band is
usually dominated by two spectral components: a soft thermal
component that probably comes from the accretion disk and a hard
power-law component. This hard component has often been associated with
the inverse-Compton scattering of thermal photons through a corona of hot
electrons, and more recently also with emission from the base of
a jet \citep{marko2003}. However, the existence of QPOs in the
radio-faint observations of GRS 1915+105 suggests that they do
not originate in a jet \citep{mille05}. Type C QPOs, and in fact many
types of QPOs in X-ray  binaries, have often been associated with
this hard spectral component---not only because the fractional
amplitude of the QPOs increases with energy, but also because the
QPOs are detected in energy bands where the contribution from the
disk is negligible. In view of these facts, it seems clear that while
the disk may determine the seed frequencies, it is likely that some other
geometry is also involved in producing the non-thermal component and
the QPO behavior. For example, it has long been proposed that the
inner-most section of a thin accretion disk may be very hot, optically
thin, and thus radiatively inefficient \citep{thorn75}. We assume this
medium of hot gas is closely 
aligned with the inner edge of the accretion disk, which in our model
takes the form of a geodesically precessing ring.

The GRS 1915+105 observations of \citet{mille05} can be explained well
with the precessing ring model, matching the observed amplitudes of
the fundamental and harmonic peaks of the QPO power spectrum. The
rms variations of $13-15\%$, coupled with an independently measured
inclination of $\iBH = i_{\rm bin} = 66^\circ$ \citep{fende99}, suggest a
tilt angle of $\Delta\theta \approx 15-20^\circ$. The spectral line
measurements suggest a moderate to
high spin with $a/M \gtrsim 0.5$ so that $\dE \lesssim 1.5$ keV,
but the {\it RXTE} spectral data does not
have enough energy resolution to rule out lower values of $a/M$. In
short, our ring model agrees quite well with the light curve
modulations, giving good estimates for the overall geometry of the
system, yet the poor spectral data still cannot sufficiently constrain
the black hole spin $a/M$.

Until recently, only time-averaged iron emission line profiles, and
lines drawn from flux windows during aperiodic variability, had been
studied with moderate and high resolution spectrometers. The recently
launched {\it Suzaku} and future
missions such as {\it Constellation-X} and {\it XEUS}
promise high spectral resolution along with high effective area.  This is
exactly the combination required to rigorously test the predictions of
our precessing ring QPO model.

{\it Suzaku} X-ray Imaging Spectrometer (XIS) observations of
GRS~1915$+$105 for $\sim$100~ksec in states similar to those considered
by \citet{mille05} should achieve the same flux sensitivity as
obtained with {\it RXTE} but with a spectral resolution ten times
higher ($\sim$100~eV versus $\sim$1~keV at 6~keV).  This may be
sufficient to clearly detect a third peak in the iron emission line
profile using the same QPO-phase-resolved technique.
Depending on their final configurations, {\it
Constellation-X} and {\it XEUS} may be able to achieve the same flux
sensitivity as the \citet{mille05} result in only a few ksec
(assuming similar QPO amplitudes and frequencies), with an energy resolution
of a few eV at 6~keV.  With such observatories, a third peak in the
iron emission line profile, if present, should be clearly resolved.
Indeed, with {\it Constellation-X} and {\it
XEUS}, it may be possible to detect a modulation of the iron line flux
and profile in as few as 100 QPO cycles, which would enable detailed
measurements of the inner disk ring evolution with time.

While this paper has focused primarily on stellar-mass black holes,
and a single source in particular, the ring model could easily be
applied to AGN observations as well. In fact, the much longer time
scales associated with AGN would greatly improve our spectral
resolution by using ``slower'' observatories like {\it Chandra} and
{\it XMM-Newton}. Since the spin of an AGN is thought to come
primarily from accretion, it is
likely the accretion disk and black hole spin axes are closely
aligned, potentially giving a small tilt $\Delta\theta$ and thus little
variation in the light curve. Also, the massive accretion disk around
an AGN is almost certainly optically thick, decreasing the relative
effects of gravitational lensing (photons cannot pass through the disk
and form Einstein rings). However, due to the larger signal-to-noise ratio
for AGN timing observations, no QPOs have yet been unambiguously
identified in AGN, making a direct comparison to galactic black holes
more difficult. 

\vspace{0.25cm}\noindent We thank Chris Reynolds for helpful
discussions and Cole Miller for his extensive and
constructive comments to an early version of the manuscript.
JDS is grateful to Ed Bertschinger for his continued
insights and encouragement. Support comes from NASA grant NAG5-13306.

\begin{table}
\caption{\label{tableone} QPO-rms measurements for seven black hole
X-ray binaries with different inclinations}
\vspace{0.2cm}
\begin{tabular}{cccccc}

\hline
\hline
Source & Inclination & QPO frequency & QPO rms & Ref.    & {\it RXTE} \\
       & (degrees)   & (Hz)          & (\%)    &         & Obs. ID \\
\hline
4U 1543--47     &  20.7$\pm$0.5 &  4.49$\pm$0.05 & 6.5$\pm$0.5  & 1 &
70128-01-01-00 \smallskip \\ 
GX 339-4        &  15--30       &  4.78$\pm$0.09 & 7.2$\pm$1.2  & 2 &
70110-01-95-00 \\
                &               &  1.24$\pm$0.01 & 5.9$\pm$0.5  &   &
70109-01-06-00 \smallskip \\ 
XTE J1650--500  &  $>$50$\pm$3  &  4.67$\pm$0.03 & 5.8$\pm$0.3  & 3 &
60113-01-12-03 \\ 
                &               &  1.31$\pm$0.14 & 5.0$^{+0.6}_{-0.4}$
&   & 60113-01-05-00 \smallskip \\ 
H1743--322      &  60--70       &  4.75$\pm$0.01 & 9.98$\pm$0.13& 4 &
80146-01-03-00 \smallskip \\ 
GRS 1915+105    &  66$\pm$2     &  1.04$\pm$0.01 & 11.3$\pm$0.1 & 5 &
20402-01-50-01 \smallskip \\ 
GRO J1655--40   &  70.2$\pm$1.2 &  1.35$\pm$0.01 & 19.3$\pm$0.3 & 6 &
91502-01-01-13 \smallskip \\ 
XTE J1550--564  &  72$\pm$5     &  4.73$\pm$0.01 & 11.7$\pm$0.2 & 7 &
30191-01-28-01 \\ 
                &               &  1.04$\pm$0.01 & 15.8$\pm$0.3 &   &
30188-06-01-03 \\ 
                &               &  1.54$\pm$0.01 & 15.8$\pm$0.3 &   &
30188-06-04-00 \\ 
\hline

\end{tabular} \\
{\bf References: 1.} \citet{orosz03} 
{\bf 2.} \citet{wu01} 
{\bf 3.} \citet{orosz04} \\
{\bf 4.} \citet{homan05} 
{\bf 5.} \citet{fende99}
{\bf 6.} \citet{green01} \\
{\bf 7.} \citet{orosz02} 
\end{table}

\newpage

\end{document}